\documentclass{article}
\usepackage{graphics}
\usepackage[dvipdfmx]{graphicx}
\usepackage{multicol}
\usepackage{enumerate}
\begin{document}
\title{Numerical study on the emergence of anisotropy in artificial flocks: \\
A BOIDS modeling and simulations of empirical findings}
\author{Motohiro Makiguchi and Jun-ichi Inoue \\
\mbox{}\\
Complex Systems Engineering, 
Graduate School of Information Science and 
Technology \\ 
Hokkaido University, N14-W9, Kita-Ku, Sapporo 060-0814, Japan}
\date{November 29, 2009}
\maketitle
\begin{abstract}
In real flocks, it was revealed that the angular density of nearest
neighbors shows a strong {\it anisotropic structure} of individuals by very
recent extensive field studies by Ballerini et al [{\it Proceedings of the
National Academy of Sciences USA} {\bf 105},  pp.1232-1237 (2008)]. In
this paper, we show that this empirical evidence in real flocks, namely,
the structure of anisotropy also emerges in an artificial flock
simulation based on the {\it BOIDS} by Reynolds [{\it Computer Graphics} 
{\bf 21}, pp.25-34 (1987)]. We numerically find that appropriate combinations of
the weights for just only three essential factors of the BOIDS, namely,
`Cohesion', `Alignment' and `Separation' lead to a strong anisotropy in
the flock. This result seems to be highly counter-intuitive and also provides a
justification of the hypothesis that the anisotropy emerges as a result
of self-organization of interacting intelligent agents (birds for instance). To quantify
the anisotropy, we evaluate a useful statistics (a kind of {\it order parameters} in statistical physics), 
that is to say, the so-called $\gamma$-value defined as an inner product between the vector
in the direction of the lowest angular density of flocks and the vector
in the direction of the moving of the flock. Our results concerning the
emergence of the anisotropy through the $\gamma$-value might enable us to
judge whether an arbitrary flock simulation seems to be {\it realistic} or not.
\end{abstract}
\mbox{}\\
{\bf keywords}: Self-organization, Anisotropy, BOIDS, Swarm Intelligence Simulation, Collective behaviour \\
\section{Introduction}
Collective behaviour of interacting {\it intelligent} agents such as birds, insects or fishes 
shows highly non-trivial properties and sometimes it seems to be 
quite counter-intuitive \cite{Modeling}. As well-known, many-body systems having a lot of {\it non-intelligent} 
elements, for instance, spins (tiny magnets in atomic scale length), 
particles, random-walkers etc. also show a collective behaviour like  
a critical phenomenon of order-disorder phase transitions with 
`spontaneous symmetry breaking' in spatial structures of the system.  
Up to now, a huge number of numerical studies in order to figure it out have been done 
by theoretical physicists and mathematicians \cite{Landau}. 
They attempted to describe these phenomena by using 
some probabilistic models and revealed the  `universality class' 
of the critical phenomena by solving the problem with the assistance 
of computer simulations. 
Of course, the validity of the studies should be checked by comparing 
the numerical results with the experimental findings. 
If their results disagree with the empirical data, the models they used 
should be thrown away or should be modified appropriately. 

On the other hand, for the mathematical modeling of many-body systems 
having interacting {\it intelligent} agents (animals), we also use some 
probabilistic models, however, it is very difficult for us to 
evaluate the modeling and also very hard to judge whether it looks like {\it realistic} or not due to 
a lack of enough empirical data to be compared. 
 
 One of the key factors for such non-trivial collective behaviour 
 of both {\it non-intelligent} and {\it intelligent} agents  
 is obviously a `competition' between several different (and for most of the cases, 
 these are incompatible) effects. 
 For instance,  the Ising model as an example of collective behaviour 
 of {\it non-intelingent} agents exhibits an order-disorder phase transition \cite{Landau} 
 by competition between the ferromagnetic interactions between Ising spins 
 (`energy minimization') 
 and thermal fluctuation (`entropy maximization') by controlling the temperature of the system. 
 On the other hand,   
 as a simplest and effective algorithm in computer simulations 
 for flocks of {\it intellingent} agents, say,  animals such as starlings, the so-called BOIDS founded by Reynolds 
 \cite{Reynolds,BOIDS}
 has been widely used not only in the field of computer graphics but also in 
 various other research fields  including ethology, physics, 
 control theory, economics, and so on. 
 The BOIDS simulates the collective behaviour of animal flocks 
 by taking into account only a few simple rules for each 
 interacting {\it intelligent} agent.
  
However, there are few studies to compare the results 
of the BOIDS simulations with the empirical data. 
Therefore, the following essential and interesting queries still have been left unsolved; 
\begin{itemize}
\item
What is a criterion to determine to what extent 
the flocks seem to be {\it realistic}? 
\item
Is there any quantity (statistics) to measure the {\it quality} of 
the artificial flocks? 
\end{itemize}
From the view point of `engineering', the above queries are (in some sense)  
not essential because their main goal is to construct a useful algorithm based on the collective behaviour of agents. 
However, from the natural science view points, 
the difference between empirical evidence and the result of the 
simulation is the most important issue and the consistency is a guide to judge the validity of the computer modeling and 
simulation. 

Recently, Ballerini {\it at al} \cite{Ballerini}
succeeded in obtaining the data for such collective  
animal behaviour, namely, empirical data of 
starling flocks containing up to a few thousands members. 
They also pointed out 
that the angular density of the nearest neighbors in the 
flocks is not uniform but 
apparently biased (it is weaken) along the direction of the flock's motion. 

With their empirical findings in mind, in this paper, we examine the possibility of the BOIDS simulations 
to reproduce this {\it anisotropy} and we also  
investigate numerically the condition on which the anisotropy emerges. 

This paper is organized as follows. 
In the next section, we explain the empirical findings 
by Ballerini {\it et al} \cite{Ballerini,Cavagna} and introduce a key concept 
{\it anisotropy} and a relevant quantity $\gamma$-value. 
In section 3,  the BOIDS modeling and 
setting of essential parameters in our simulations are 
explicitly explained. 
The results are reported in section 4. 
The last section provides concluding remarks. 
\section{Empirical findings by Ballerini et al}
In this section, we briefly review the measurement of the realistic flocks 
and the evaluation of the empirical data 
by Ballerini {\it et al} \cite{Ballerini}. 
They measured each bird's position 
 in the flocks  of starling (\textit{Sturnus vulgaris}) in three dimension.
To get such 3D data, they used `Stereo Matching'  which reconstructs 3D-object
 from a set of stereo photographs.
 \subsection{Anisotropy}
From these data, they calculated  the angular density of the nearest neighbours 
in the flock.
They measured the angles ($\phi$, $\alpha$), 
where $\phi$ stands for the `latitude' of the nearest neighbour for 
each bird measured from the direction of the motion of the flock,  
whereas $\alpha$ denotes `longitude' which specifies the 
position of the nearest neighbour for each bird around the direction of flock's motion, 
for all individuals in the flock 
and made the 2D-map of angular density distribution 
using the so-called `Mollweide projection'. 
Their figure clealy shows that the density is not uniform but 
obviously  biased.  For instance, we find from the figure that 
the dinsity around $\phi \simeq 0$ and 
$\alpha \simeq 0^{\rm o}, \pm 180^{\rm o}$ are extremely  low 
in comparison with the density in the other directions. 
The property of the biased distribution due to the absence of 
the birds along the direction of the flock's motion is referred to as {\it anisotropy} \cite{Ballerini}.
The main goal of this paper is to reveal numerically that 
the artificial flock by the BOIDS 
exhibits the anisotropy as the realistic flock shows \cite{Ballerini}. 
To quantify the degree of the anisotropy, we use a useful 
statistics (a kind of `order parameters' in the research field of statistical mechanics) 
introduced in the following subsections. 
\subsection{The $\gamma$-value: An order parameter to detect `spatial symmetry breaking'}
Ballerini {\it et al} also introduced a useful indicator, 
what we call  $\gamma$-value.
The $\gamma$-value is calculated according to the following recipe.
Let $\mbox{\boldmath $u$}_i^{(n)}$ be an unit vector pointing in the direction of
 the $n^{th}$ -nearest neighbour of the bird $i$ 
 and let us define the projection matrix 
$\mbox{\boldmath $M$}^{(n)}$ in terms of 
the $\mbox{\boldmath $u$}_i^{(n)}$ as follows. 
\begin{eqnarray*}
 \mbox{\boldmath $u$}_i^{(n)}  =  \left(
	\begin{array}{c}
		{u_i}_x^{(n)}\\
		{u_i}_y^{(n)}\\
		{u_i}_z^{(n)}
	\end{array} \right),
	(\mbox{\boldmath $M$}^{(n)})_{\alpha \beta}  = \frac{1}{N}\sum_{i=1}^{N}
	{(\mbox{\boldmath $u$}_i^{(n)})^\alpha
	(\mbox{\boldmath $u$}_i^{(n)})^\beta}\,\,\, (\alpha,\beta=x, y, z)
\end{eqnarray*}
where $N$ is the number of birds in the flock. 
Then, the $\gamma$-value is given by
\begin{eqnarray}
\gamma & = & \langle {(\mbox{\boldmath $W$}^{(n)} \cdot 
\mbox{\boldmath $V$})}^2 \rangle 
\label{eq:def_gamma}
\end{eqnarray}
where $\mbox{\boldmath $W$}^{(n)}$ 
denotes the normalized eigenvector corresponding to 
the smallest eigenvalue of the projection matrix $\mbox{\boldmath $M$}^{(n)}$.
From the definition, the $\mbox{\boldmath $W$}^{(n)}$ coincides with the direction of 
the lowest density in the flock.
The vector $\mbox{\boldmath $V$}$ appearing in the 
equation (\ref{eq:def_gamma}) means the unit  vector of flock's motion. 
The bracket $\langle \cdots \rangle$ means the average over 
the ensembles of the flocks.  
The $\gamma$-value 
for the uniform distribution 
of the position $(\varphi,\theta)$ 
for a given vector $\mbox{\boldmath $V$}$, 
namely, the $\gamma$ for $\rho (\varphi, \theta) =(4\pi)^{-1}$ is 
easily calculated as 
\begin{eqnarray}
\gamma_{isotropy} & = & 
\int_{0}^{2\pi}
d\varphi 
\int_{-\pi}^{\pi} d\theta 
\rho (\varphi,\theta) 
\sin \theta \cos^{2}\theta \nonumber =\frac{1}{3}
\end{eqnarray}
where we used $
{(\mbox{\boldmath $W$}^{(n)} \cdot 
\mbox{\boldmath $V$})}^2 = \cos^{2} \theta$. 
Therefore,  the distribution of the $n^{th}$-nearest neighbours has  
anisotropic structure when  the $\gamma$-value is larger than $1/3$,
 namely the condition for the emergence of the anisotropy is 
 explicitly written by 
\begin{eqnarray}
	\gamma & > & \gamma_{isotropy}=\frac{1}{3}.
\end{eqnarray}
By measuring this $\gamma$-value for artificial flock simulations,
one can show that the anisotropy also emerges in computer simulations.   
To put it into other words, 
the system of flocks is spatially `symmetric' for $\gamma = \gamma_{isotropy}$, whereas 
the symmetry is `spontaneously'  broken for $\gamma > \gamma_{isotropy}$. 
This `spontaneous symmetry breaking' is nothing but the emergence of anisotropy. 

In the following sections, 
we carry out the BOIDS simulations and 
evaluate the anisotropy by the $\gamma$-value. 
Then, we find that the `spontaneous symmetry breaking'  mentioned above 
actually takes place by controlling the essential parameters appearing in the BOIDS. 
\section{The BOIDS modeling and simulations}
To make flock simulations in computer,
 we use the so-called BOIDS which was originally designed by Reynolds \cite{Reynolds}.
The BOIDS is one of the well-known mathematical (probabilistic) models in the research fields of CG and animation. 
Actually, the BOIDS can simulate very complicated animal flocks or schools 
although it consists of just only three simple interactions for each agent in the aggregation: 
 \begin{enumerate}
 \setlength{\topsep}{0pt}
 \setlength{\partopsep}{0pt}
 \setlength{\itemsep}{0pt}
 \setlength{\parsep}{0pt}
 \setlength{\parskip}{1pt plus 1pt minus 1pt}
 \item[(c)]{\bf Cohesion}: Making a vector of each agent's position toward the average position of local flock mates.
 \item[(a)]{\bf Alignment}: Making a vector of each agent's position towards the average heading of local flock mates and 
 keeping  the velocity of each agent the average value of flock mates.
 \item[(s)]{\bf Separation}: Controlling the vector of each agent  to avoid the collision with the other local flock mates.
\end{enumerate}
It is important for us to bear in mind that 
`local flock mates' mentioned above denotes the neighbours within the range of views for each agent.
Each agent decides her (or his) next migration by compounding these three interactions. 
\subsection{On the setting of essential parameters in BOIDS simulations}
\begin{figure}[htpb]
\begin{center}
\includegraphics[bb= 0 0 299 300,width=5cm]{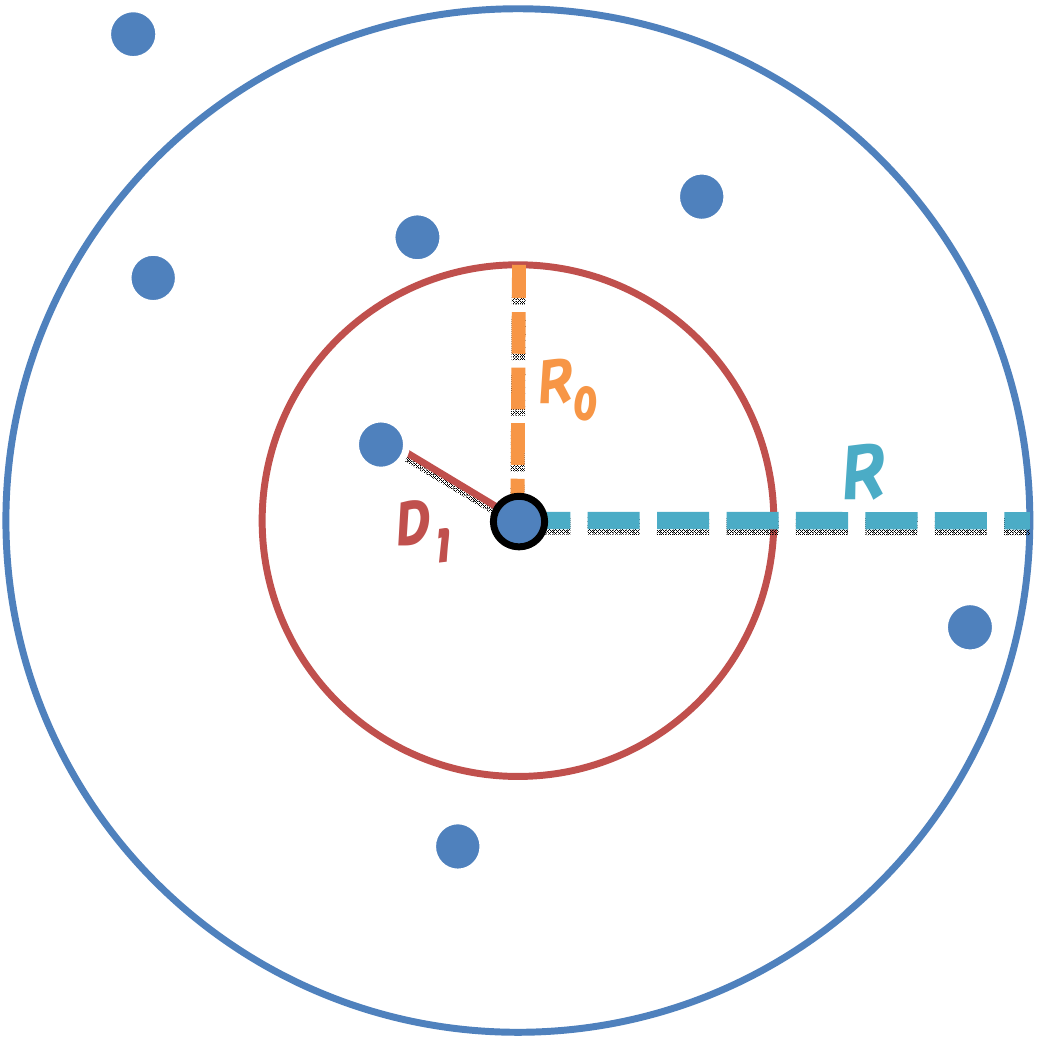}
\end{center}
\caption{\footnotesize
Range of View ($R$) and Separation($R_{0}$).}
\label{fig:range}
\end{figure}

In our BOIDS simulations, each agent is defined as a mass point 
 and specified by a set of 
 3D-coordinate $(x,y,z)$, an unit vector of motion,
 and the speed.
We define each agent's view as
 a sphere with a radius $R$ without any blind corner. 
 We also define the `separation sphere'  with a radius $R_0$  and 
 the distance between the nearest neighbours is specified by the length  $D_{1}$ (see Figure \ref{fig:range}).
The interaction of `Separation' is switched on 
if and only if the $D_{1}$ is smaller than the $R_{0}$.
Some other essential parameters appearing in our BOIDS simulations and 
the setup are also explicitly given as follows. 
\begin{itemize}
 \setlength{\topsep}{0pt}
 \setlength{\partopsep}{0pt}
 \setlength{\itemsep}{0pt}
 \setlength{\parsep}{0pt}
 \setlength{\parskip}{1pt plus 1pt minus 1pt}
 \item {\bf Field of simulations: } Three dimensional open space without any gravity or 
 	any air resistance.  Moreover, there is no wall and no ground surface.
 \item {\bf The number of agents in the flock:} The system size of simulations is $N=100$
 \item {\bf Initial condition on the speed of each agent:} $1< speed <8$. 
 \item {\bf Initial condition on the location of each agent:} All agents are distributed in a sphere with radius  $L=180$. 
 \item {\bf The shape and the range of each agent's  view:} A sphere with radius $R=200$.
 \item {\bf The shape and the range of separation:} A sphere with radius $R_0=20$. 
\end{itemize}
For the above setting of the parameters, 
we shall implement two types of programming codes.
One is a programing code for a single simulation (`SS'  for short),
 and another code is for multiple simulations (`MS' for short).
The SS runs in the GUI (graphical user interface) and it shows us a shape of the flock in real time, 
whereas 
the MS enables us to carry out a number of simulations 
with different initial conditions.
\subsection{Typical four aggregations}
By controling the three essential interactions, namely, 
`Cohesion', `Alignment' and `Separation'  mentioned above,
 we obtain four different aggregations having different collective behaviours.
Each behavour of the aggregations is monitored (observed) by the SS.
To specify the aggregation process of the flocks, we define the 
update rule of the vector of movement 
for each agent by 
\begin{eqnarray}
\mbox{\boldmath $v$} (t+1) & = & 
\frac{
P_{0}\mbox{\boldmath $V$}_{\rm Cohesion} (t) + 
P_{1}\mbox{\boldmath $V$}_{\rm Alignment} (t) + 
P_{2}\mbox{\boldmath $V$}_{\rm Separation} (t)  + 
P_{3}\mbox{\boldmath $v$} (t)}
{|P_{0}\mbox{\boldmath $V$}_{\rm Cohesion} (t) + 
P_{1}\mbox{\boldmath $V$}_{\rm Alignment} (t) + 
P_{2}\mbox{\boldmath $V$}_{\rm Separation} (t)  + 
P_{3}\mbox{\boldmath $v$} (t)|}
\end{eqnarray}
where 
$t$ means the time step of the update 
and $|\mbox{\boldmath $A$}|$ 
denotes the $L_{1}$-norm of a vector $\mbox{\boldmath $A$}$. 
 $\mbox{\boldmath $V$}_{\rm Cohesion}$ 
is a vector pointing to the center of mass from 
each agent's position. 
$\mbox{\boldmath $V$}_{\rm Alignment}$ 
denotes a vector to be obtained 
by averaging over the velocities of all agents. 
$\mbox{\boldmath $V$}_{\rm Separation}$ means 
a vector pointing to the direction 
of the movement of each agent to be separated from 
her (or his) nearest neighbouring mate. 
Therefore, 
the aggregation of the flock 
is completely specified by the weights of 
the above vectors, 
namely, 
$\mbox{\boldmath $P$} \equiv (P_{0},P_{1},P_{2},P_{3})$. 
Among all possible combinations of these weights 
\mbox{\boldmath $P$},  we shall pick up typical four cases.  Each property and the shape of each aggregation 
are explained as follows. 
\begin{itemize}
 \item{{\it Case 1} (\bf Crowded Aggregation)}: 
 	The aggregation obtained by controlling the interaction of `Cohesion' much stronger than  
	the others, namely, $\mbox{\boldmath $P$}=(1,0,0,1)$.
 \item{{\it Case 2} (\bf Spread Aggregation)}:
 	The aggregation obtained by controlling the interaction of `Separation' much stronger than  
	the others, namely, $\mbox{\boldmath $P$}=(0,0,1,1)$.
 \item{{\it Case 3} (\bf Synchronized Aggregation)}:
 	The aggregation obtained by controlling the interaction of `Alignment' much stronger than  
	the others, namely, $\mbox{\boldmath $P$}=(0,1,1,1)$.
 \item{{\it Case 4} (\bf Flock Aggregation)}:
 	This aggregation obtained by  adjusting every interactions appropriately, 
	namely, $\mbox{\boldmath $P$}=(1,5,1.5.1)$.
\end{itemize}
Using the MS,
we simulate each aggregation for $200$ times for different initial conditions.
In each simulation, we measure the angular distribution 
and then the value of $\gamma$ is calculated. The number of crash is also updated  
 when the coordinate of each agent is identical to (is shared with) the other agents.
We start each measurement from the time point at which the total amount of the change in every agent's speed
 is  close to $0$  through the 80 turns of the update.  
 In the next section, we explain the details of the result. 
\section{Results}
\begin{figure}[htbp]
\begin{center}
\includegraphics[bb=0 0 506 281,width=7cm]{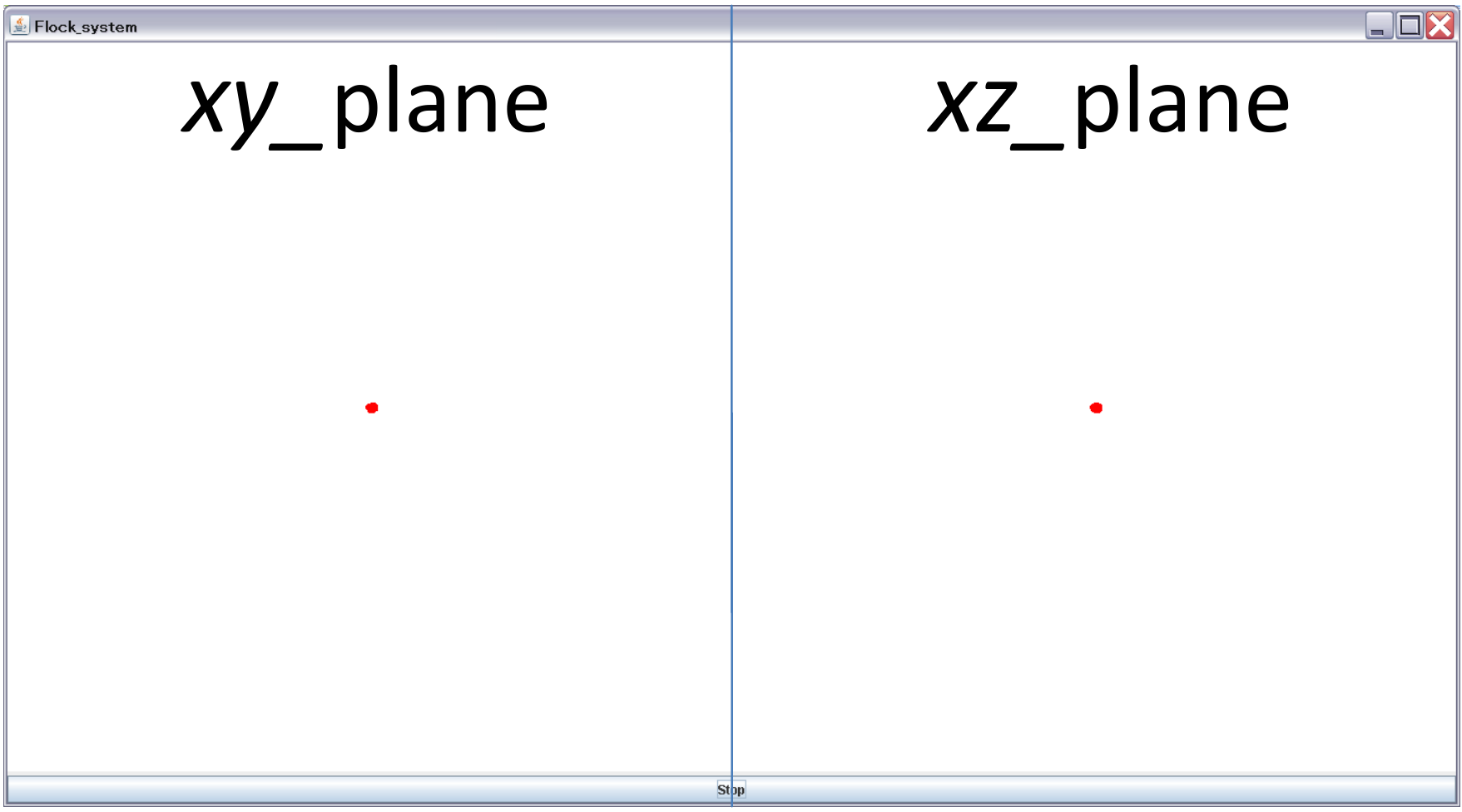} \hspace{0.1cm}
\includegraphics[bb=0 0 506 281,width=7cm]{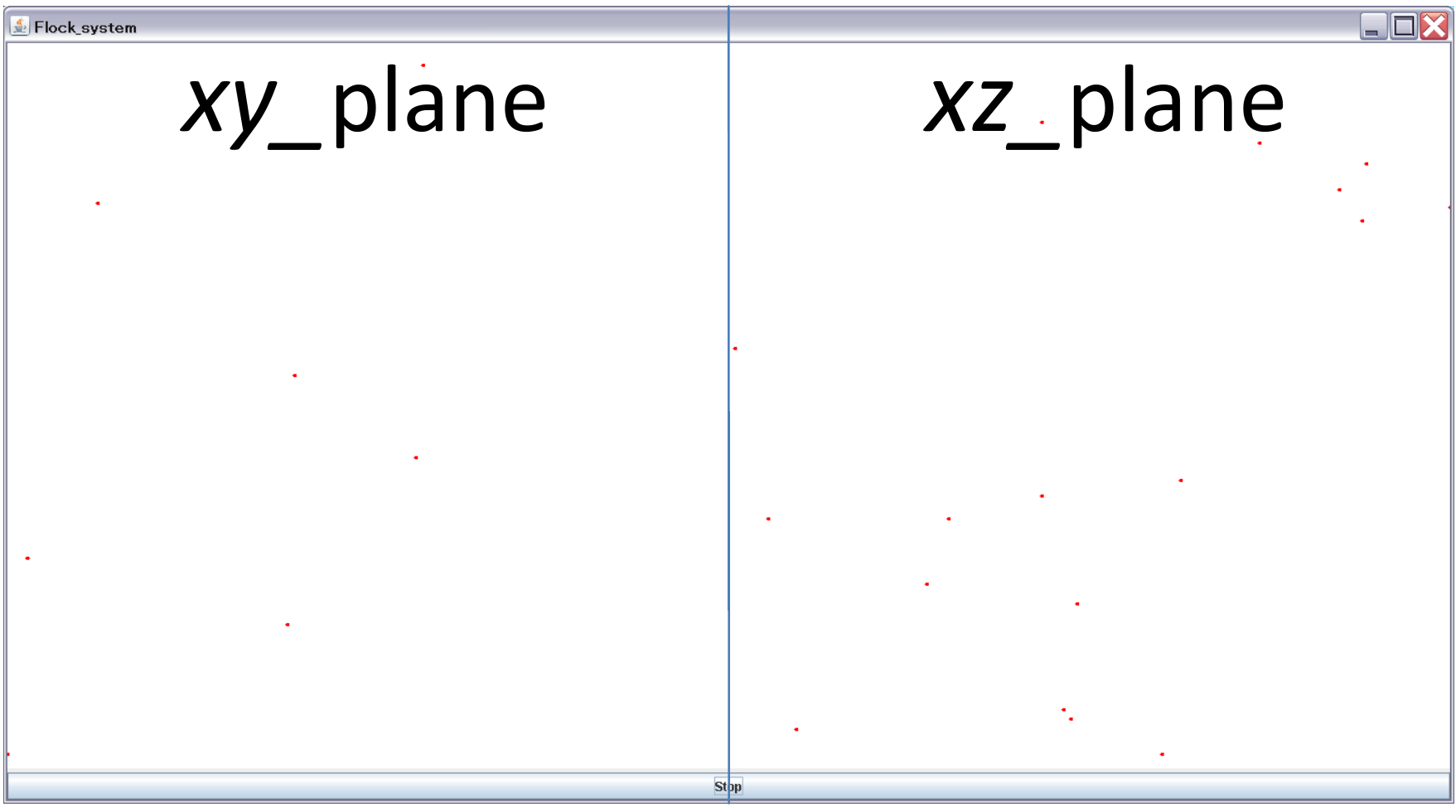} 
\mbox{}\vspace{0.1cm}
\includegraphics[bb=0 0 506 281,width=7cm]{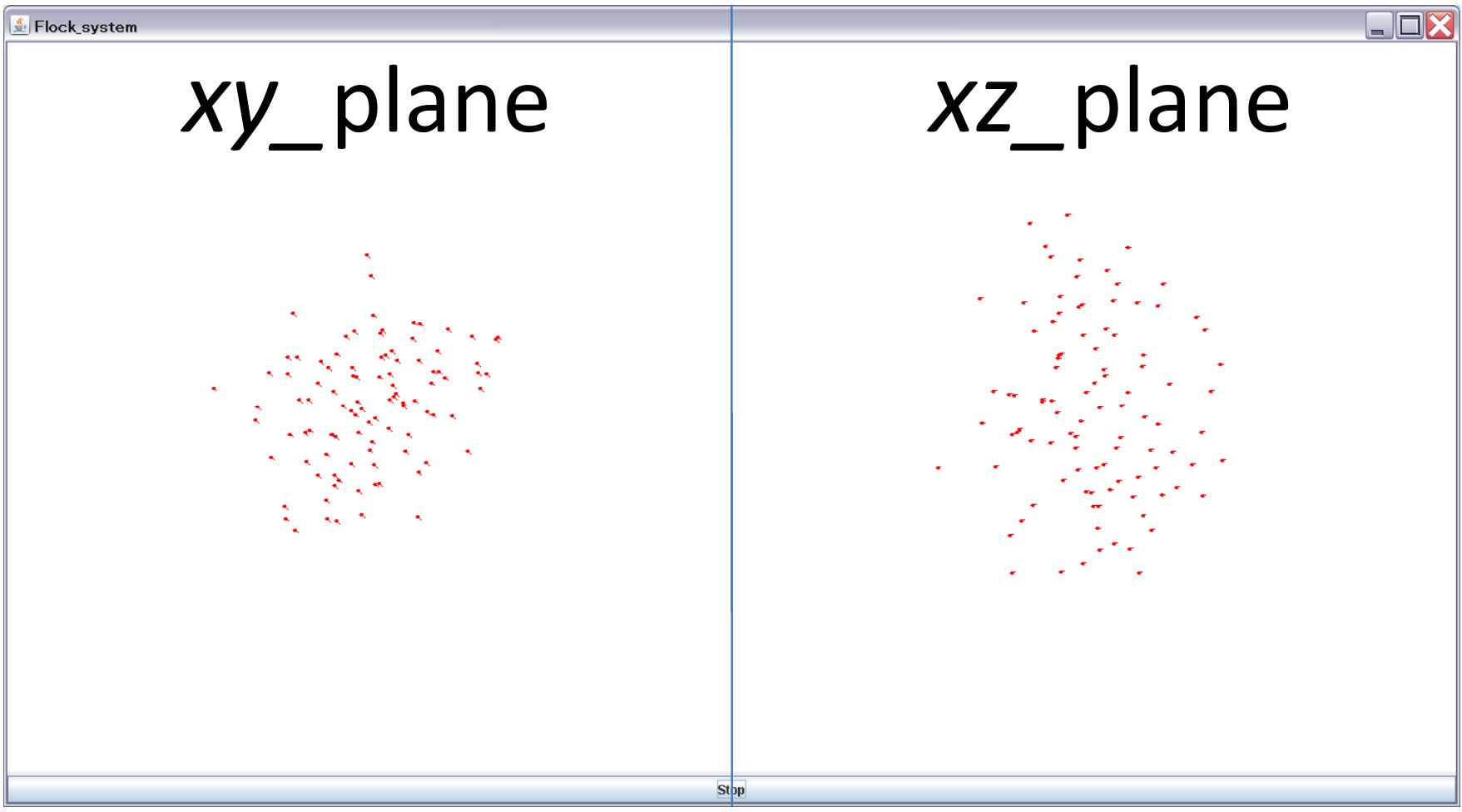} \hspace{0.1cm}
\includegraphics[bb=0 0 506 281,width=7cm]{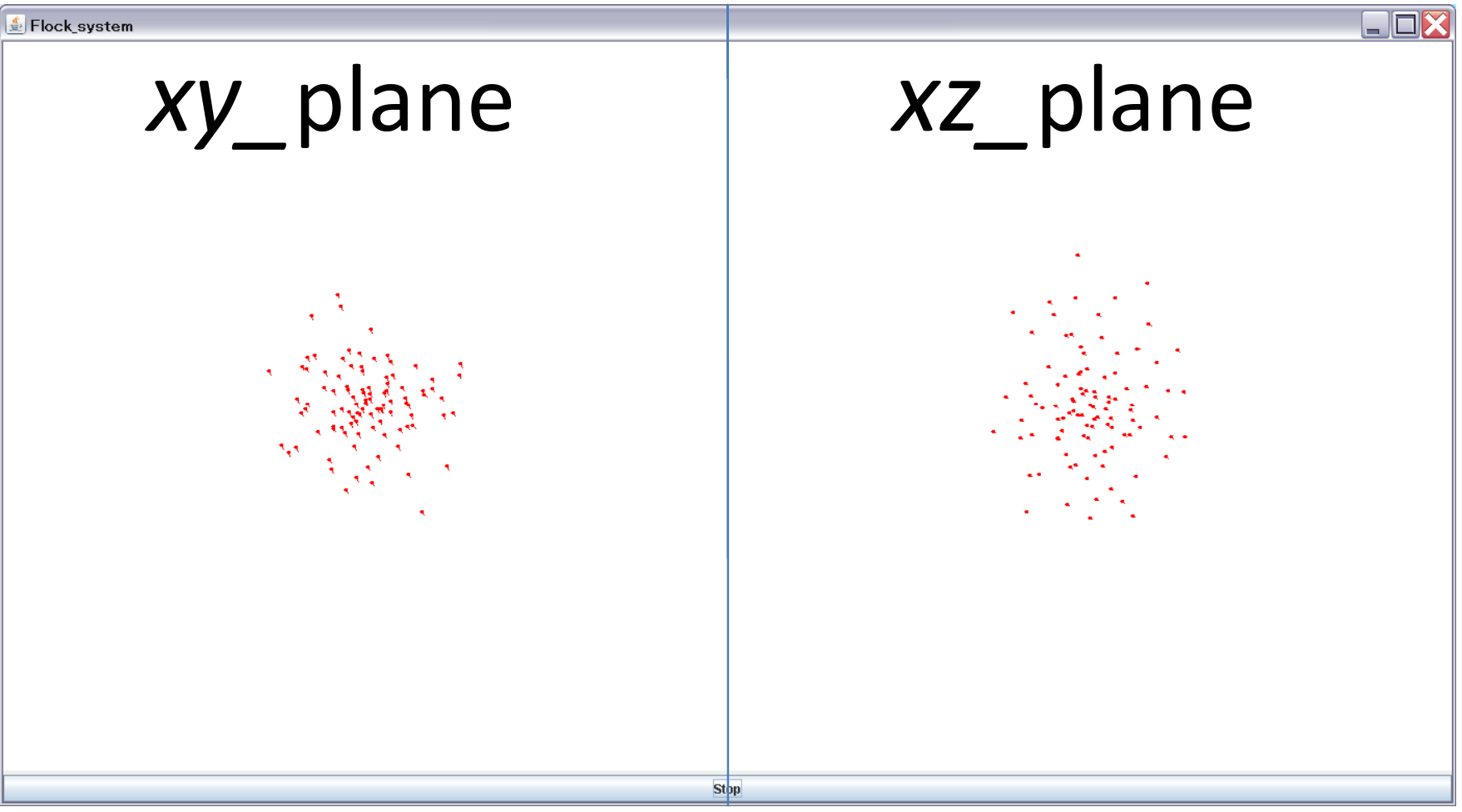}
\end{center}
\caption{\footnotesize From the upper left to the lower right, 
snapshots of the typical behaviour for `Crowded Aggregation', 
`Spread Aggregation', `Synchronized Aggregation' and 
`Flock Aggregation' (the projections to the $xy$- and $xz$-planes) are shown. }
\label{fig:fg3}
\end{figure}

\begin{figure}[htbp]
\begin{center}
\includegraphics[bb=0 0 478 247,width=7.5cm]{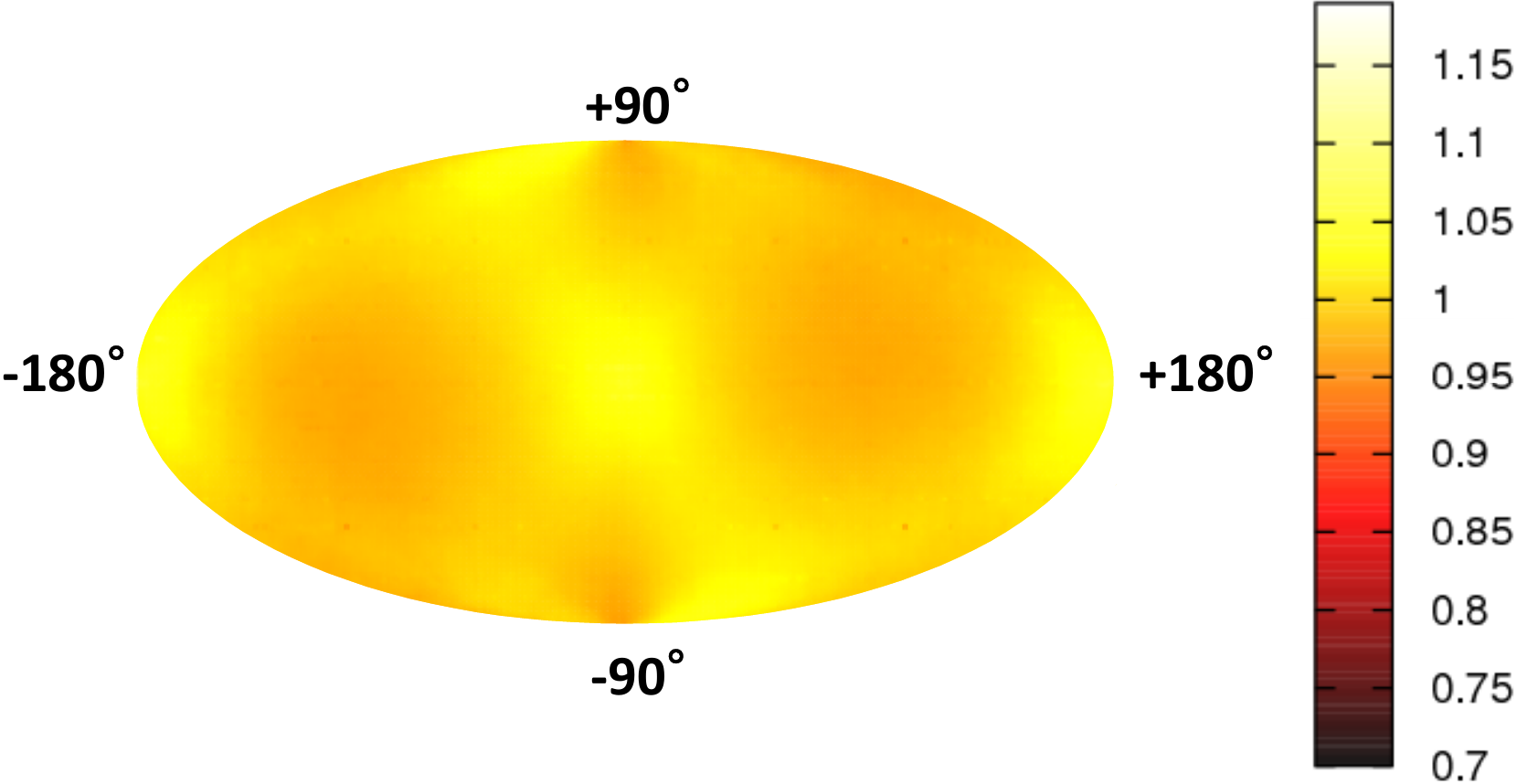} \hspace{0.1cm}
\includegraphics[bb=0 0 478 247,width=7.5cm]{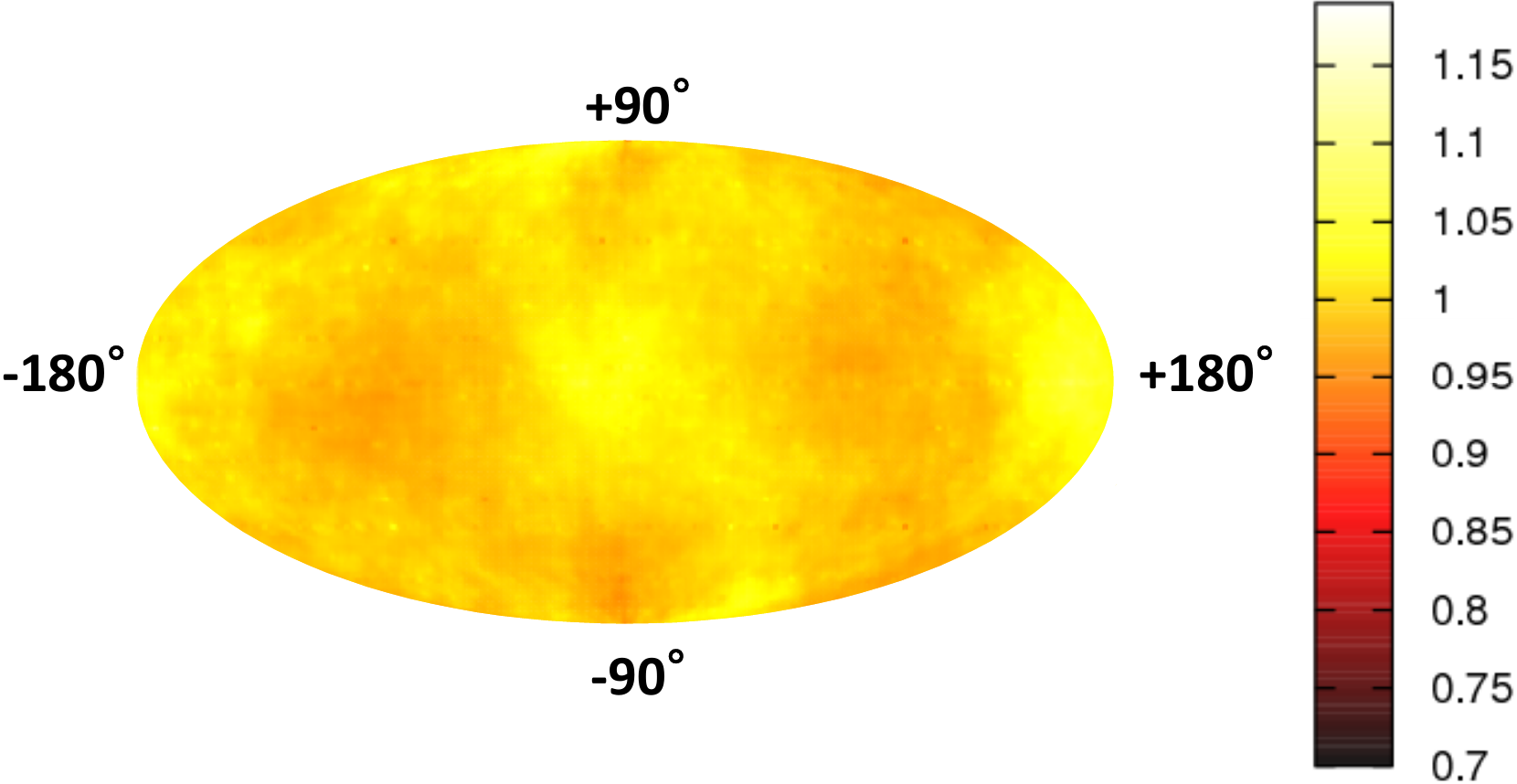}
\mbox{}\vspace{0.1cm}
\includegraphics[bb=0 0 478 247,width=7.5cm]{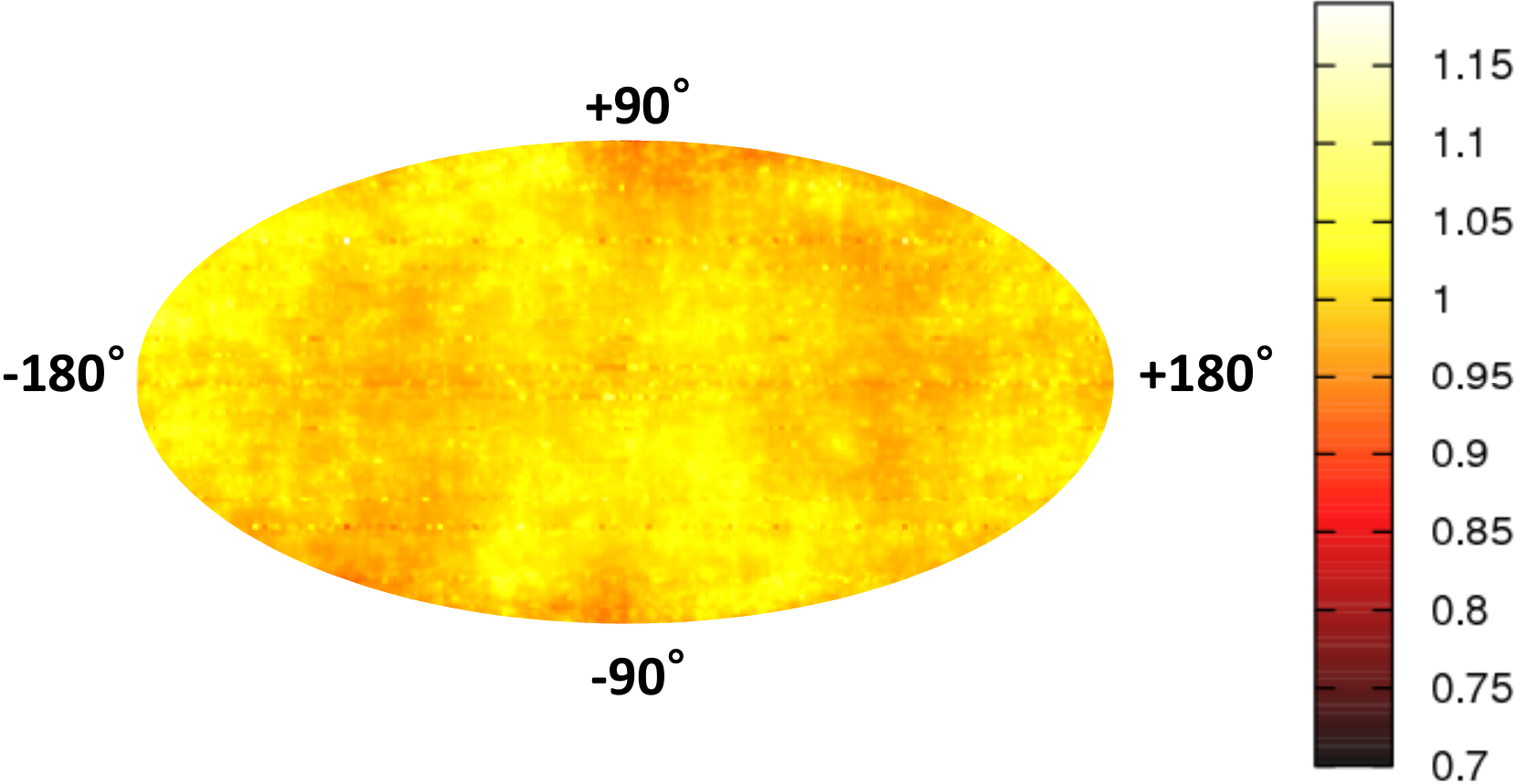} \hspace{0.1cm}
\includegraphics[bb=0 0 478 247,width=7.5cm]{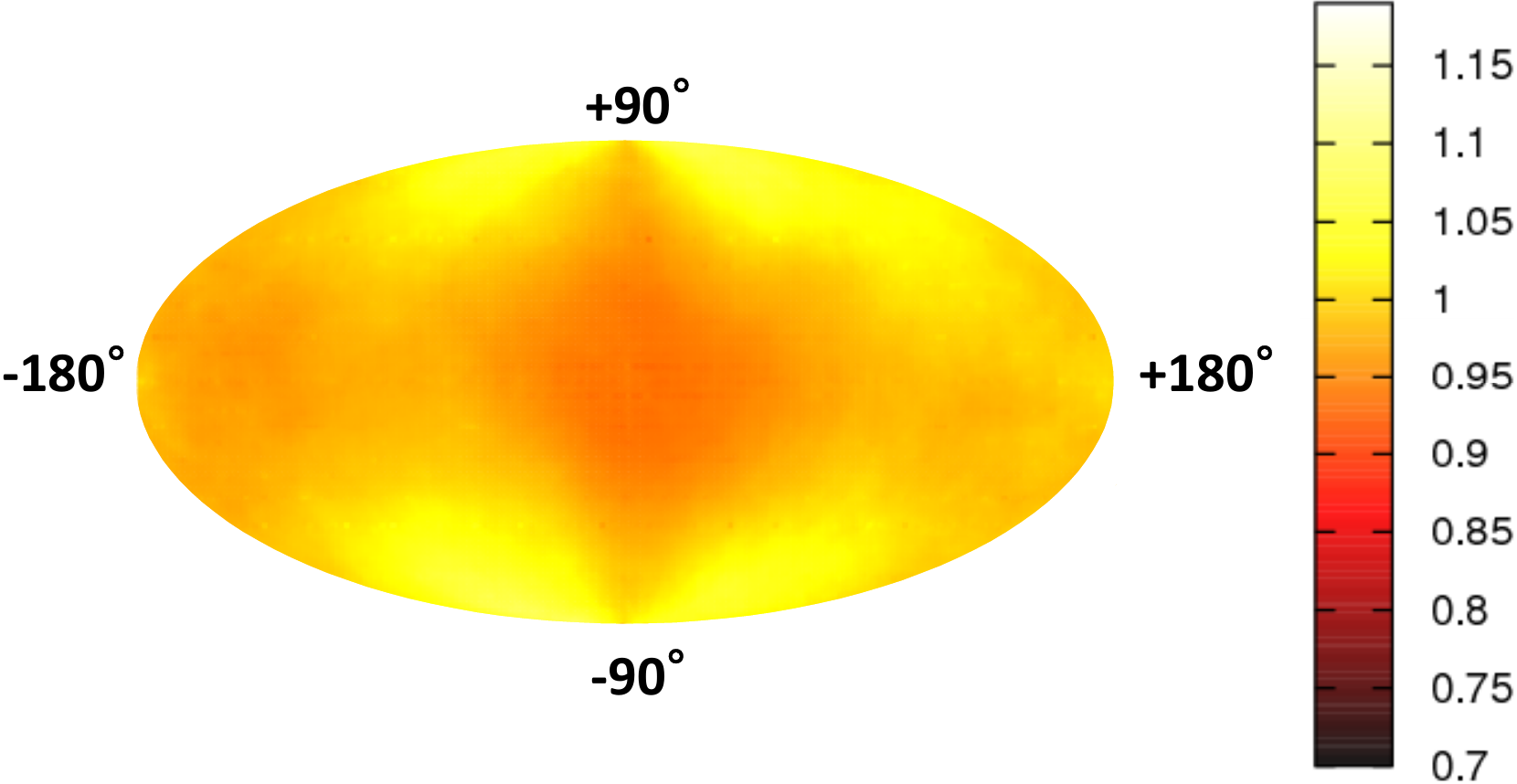}
\end{center}
\caption{\footnotesize From the upper left to the lower right, 
angular density  of the {\it Case 1} ( `Crowded Aggregation'), 
the {\it Case 2} (`Spread Aggregation'), the {\it Case 3} (`Synchronized Aggregation') 
and the {\it Case 4} (`Flock Aggregation') are shown. 
The spatial symmetry in the {\it Case 4} is apparently broken (the anisotropy emerges).
}
\label{fig:fg4}
\end{figure}

In association with the each aggregation,
 we evaluate the $\gamma$-value  with standard deviation 
 and the average number of crashes for 200 independent runs of the BOIDS simulations.
In following, we summarize the results. 
\begin{itemize}
	\item{{\it Case 1} (\bf Crowded Aggregation)}: 
		The typical behaviour of the flock and 
		the angular distribution are shown in the upper left panel in Figure \ref{fig:fg3} 
		and Figure \ref{fig:fg4}, respectively.  
		The $\gamma$-value is $0.333$ with standard deviation $0.083$ and 
		the average number of crashes is $2091.74$. 
	\item{{\it Case 2} (\bf Spread Aggregation)}: 
	         The typical behaviour of the flock and 
	         the angular distribution are shown in the upper right panel 
	         in Figure \ref{fig:fg3} and Figure \ref{fig:fg4}, respectively.  
		The $\gamma$-value is $0.315$ with standard deviation $0.258$ and 
		the average number of crashes is $0$.
	\item{{\it Case 3} (\bf Synchronized Aggregation)}: 
		The typical behaviour of the flock and
		the angular distribution are shown in the lower left panel 
		in Figure \ref{fig:fg3} and Figure \ref{fig:fg4}, respectively.  
		The $\gamma$-value is $0.300$ with standard deviation $0.281$ and 
		the average number of crashes is $0$.
 	\item{{\it Case 4} (\bf Flock Aggregation)}: 
		The typical behaviour of the flock and
		the angular distribution are shown in the lower right panel 
		in Figure \ref{fig:fg3} and Figure \ref{fig:fg4}, respectively.  
		The $\gamma$-value is $0.744$ with standard deviation $0.124$ and 
		the average number of crashes is $0$.
 		\end{itemize}
The aggregation of {\it Case 4} (`Flock Aggregation') has the highest $\gamma$-value among the four cases 
 and its angular density clearly shows a lack of nearest neighbours along the direction of flock's motion 
 leading to an anisotropy.
For all aggregations except for the {\it Case 4}, the $\gamma$-values are lower than $\gamma_{isotropy}=1/3$, 
namely, they have no anisotropy.  
From these results, we find that BOIDS computer simulations having appropriate 
weights $\mbox{\boldmath $P$}$ shows anisotropy 
structures as real flocks exhibit. It is also revealed that 
 one can evaluate to what extent an arbitrary flock simulation is close to real  flocks 
 through the $\gamma$-value.
 
\begin{figure}[ht]
\begin{center}
\includegraphics[bb=0 0 478 247,width=7.5cm]{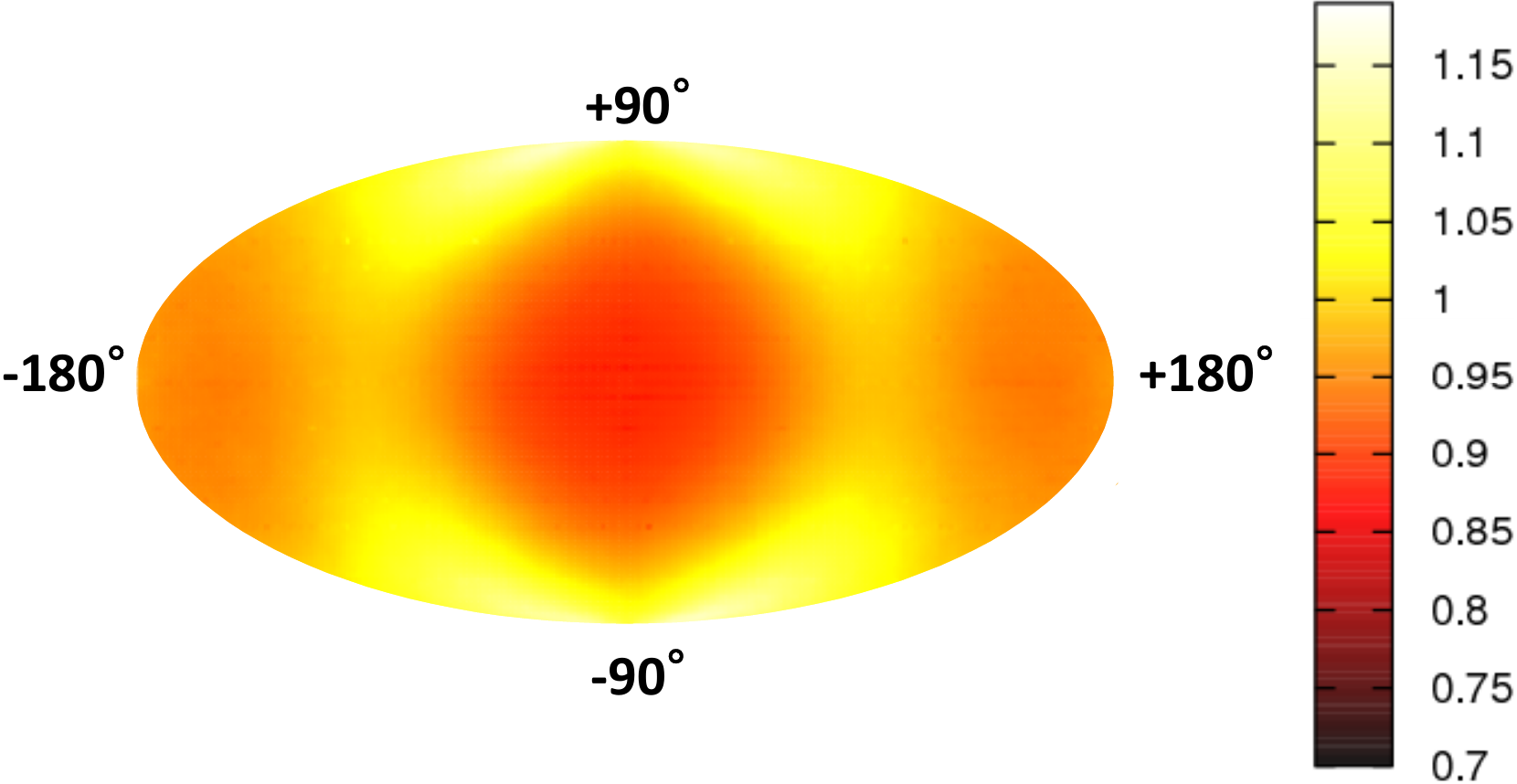}
\end{center}
\caption{\footnotesize The angular density for the aggregation ({\it Case 5}) 
having a higher $\gamma$-value than the {\it Case 4}. } 
\label{fig:fg5}
\end{figure}

From the results obtained here, we might have another question, namely, 
it is important for us to answer the question such as 
whether the aggregation having a higher $\gamma$-value than the {\it Case 4} 
seems to be more {\it realistic} than the  {\it Case 4} or not. 
To answer the question, 
 we carry out the simulations of the flock aggregation which has a higher $\gamma$-value 
 ({\it Case 5}) 
 than the {\it Case 4}. 
 The results are summarized as follows. 
\begin{itemize}
		\item{{\it Case 5} (\bf Crowded Aggregation)}: 
		The angular distribution is shown in Figure \ref{fig:fg5}.  
		The $\gamma$-value is $0.931$ with standard deviation $0.02$ and 
		the average number of crashes is $570.6$.		
\end{itemize}
Obviously,  the above aggregation has the highest $\gamma$-value 
leading to the strongest anisotropy among the five cases.
However, it is hard for us to say that it is an {\it optimal flock}   
 because the number of cashes is also the highest  ($570$ times) and 
 to make matter worse,  
 the number itself is apparently outstanding.
This result tells us that an aggregation having much stronger anisotropic structures is not always  a better flock.

The above result is reasonably accepted because the $\gamma$-value is calculated from the angular distribution
 of nearest neighbours without any concept of the {\it distance} between agents. Therefore, 
 the flock having a dense network might have highly risks of crashes more than the sparse network.  
For this reason,  in order to judge whether a given aggregation has a better flock behaviour or not,
 we should use the other criteria which take into account the distance between nearest neighbours.

\begin{figure}[ht]
\begin{center}
\includegraphics[bb=0 0 360 252,width=8cm]{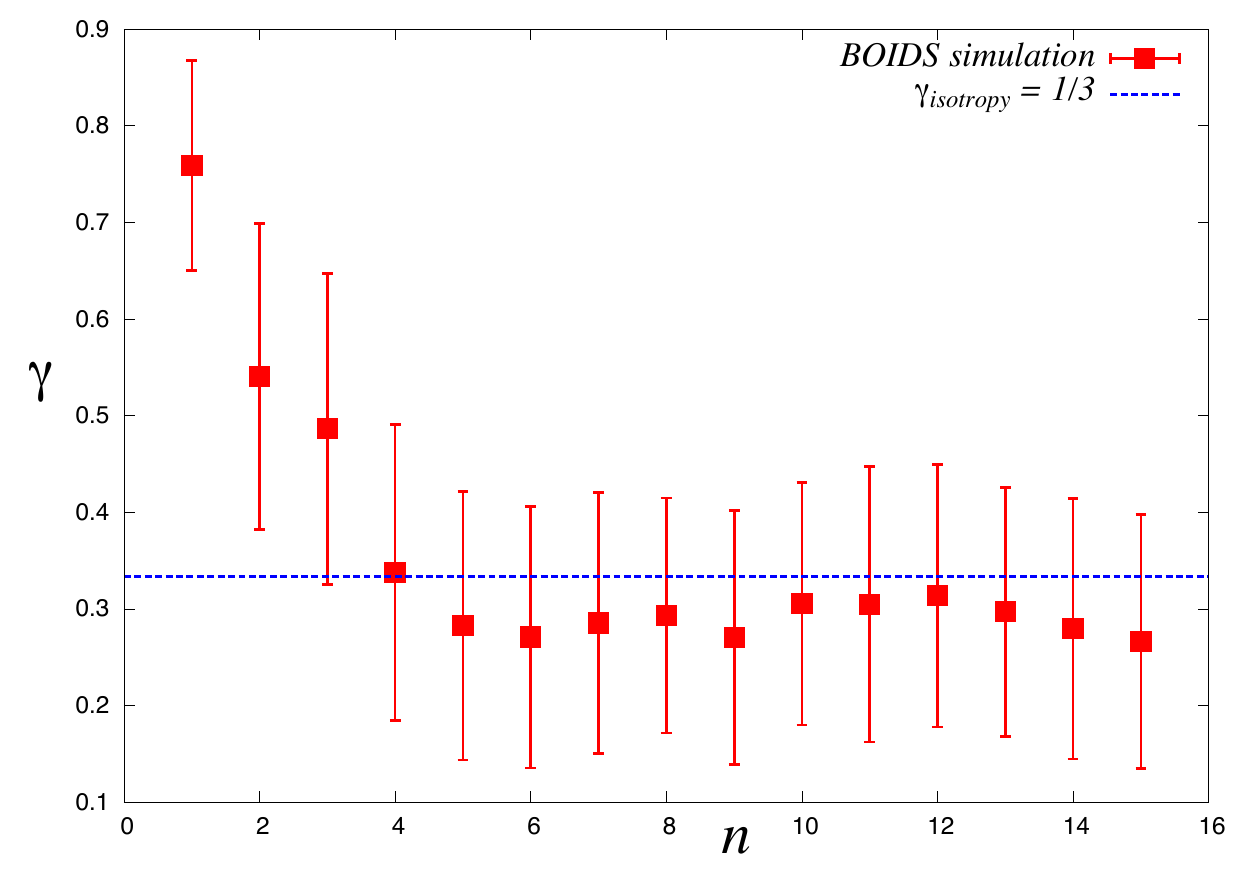}
\end{center}
\caption{\footnotesize The $\gamma$-value as a function of 
the neighbour $n$ in the real flock. }
\label{fig:fg6}
\end{figure}

Inspired by the empirical data analysis by Ballerini {\it et al} \cite{Ballerini}, 
we finally calculate the $\gamma$-value as a function of the order of the neighbour. 
The result is shown in Figure \ref{fig:fg6}. 
In this figure, $n$ denotes the order of the neighbour, 
for instance,  $n=1$ or $n=2$ means the nearest neighbour, 
 the next nearest neighbour, respectively.
The figure shows a similar behaviour to the corresponding plot in the reference \cite{Ballerini},  that is, 
the $\gamma$-values monotonically decrease as $n$ increases and they 
converge to $\gamma_{isotropy} =1/3$ beyond $n \simeq 6$. 
This result might be a justification to conclude that our BOIDS simulations having 
appropriate weight vectors $\mbox{\boldmath $P$}$ actually simulate 
a {\it realistic} flock. 
\section{Concluding remarks}
In this paper, we showed that the anisotropy observed in 
the empirical data analysis \cite{Ballerini} also emerges in our BOIDS simulations having 
appropriate weight vectors $\mbox{\boldmath $P$}$. 
From the $\gamma$-value we calculated, 
one can judge wheter an optional aggregation behaves like a {\it real flock} or not. 
The system of flocks is spatially `symmetric' for $\gamma = \gamma_{isotropy}=1/3$, whereas 
the symmetry is `spontaneously'  broken for $\gamma > \gamma_{isotropy}$. 
We found from the behaviour of `order parameter' $\gamma$ that
 this `spontaneous symmetry breaking' is nothing but the emergence of anisotropy. 
 
As well-known, there are some conjectures on the origin of the emergence of anisotropy.
For instance, the effect of bird's vision is one of the dominant hypotheses.
In fact, real starlings have lateral visual axes and each of the starlings has a blind rear sector \cite{Martin}.
If all individuals in the flock move to avoid their nearest neighbours which are hidden in their blind sectors,
the effect of the blind sector is more likely to be a factor to emerge the anisotropy of nearest neighbours in the front-rear directions. 
However, our result proved that this hypothesis is {\it NOT ALWAYS  correct}   
because agents  in our simulation have no blind sector of their views.
Nevertheless, we found that 
 our flock aggregation has an anisotropy of the nearest neighbours.
The result means that the agent's  blind as an effect of vision 
is {\it not necessarily required} to 
produce the anisotropy and much more essential factor for the anisotropy is {\it the best possible 
 combinations of  three essential interactions} in the BOIDS.
 \\
 
We hope that these results might help us to consider the relevant link between BOIDS simulations and 
empirical evidence from real world.

\section*{Acknowledgement}
We were financially supported by Grant-in-Aid Scientific Research on
Priority Areas {\it `Deepening and Expansion of Statistical Mechanical Informatics
(DEX-SMI)'} of the MEXT No. 18079001. 
One of the authors (JI) was financially supported by 
INSA (Indian National Science Academy) -  JSPS 
(Japan Society of Promotion of Science)  Bilateral Exchange Programme. 
He also thanks Saha Institute of Nuclear Physics for their warm hospitality during 
his stay in India. 

\end{document}